\documentclass[12pt]{article}

\usepackage{graphics}
\usepackage{epsfig}

\begin{document}
\hoffset -1.7cm

\setlength{\hsize}{16cm}
\setlength{\vsize}{20cm}
\renewcommand{\baselinestretch}{1.2}

\title{\bf $SU(3)_c\otimes SU(3)_L\otimes U(1)_X$ as an $SU(6)\otimes 
U(1)_X$ subgroup}
\author{{\bf R. Mart\'\i nez} \\ {\it Depto. de  F\'\i sica, 
Universidad Nacional de Colombia,} \\
{\it Bogot\'a, Colombia}\\
{\bf William A. Ponce} \\
{\it Instituto de F\'\i sica, Universidad de Antioquia,} \\
{\it A.A. 1226,
Medell\'\i n, Colombia}\\ 
{\bf Luis A. S\'anchez} \\ {\it Escuela de F\'\i
sica, Universidad Nacional de Colombia,} \\
{\it A.A. 3840, Medell\'\i n, Colombia}}

\date{}
\maketitle

\begin{abstract}

\small{An extension of the Standard Model to the  local gauge group $SU(3)_c\otimes SU(3)_L\otimes U(1)_X$ as a family independent model is presented. The mass scales, the gauge boson masses, and the masses for the spin 1/2 particles in the model are studied. The mass differences between the up and down quark sectors, between the quarks and leptons, and between the charged and neutral leptons in one family are analyzed. The existence of two Dirac neutrinos for each family, one light and one very heavy, is predicted. By using experimental results from LEP, SLC and atomic parity violation we constrain the mixing angle between the two neutral currents and the mass of the additional neutral gauge boson to be $0.00015\leq\sin\theta\leq 0$ and $1.5 \mbox{TeV}\leq M_{Z_2}$ at $95 \%$ CL. }
\end{abstract}


\section{Introduction} 

In spite of the remarkable experimental success of our leading theory 
of fundamental interactions, the so called Standard Model (SM) based 
on the local gauge group $SU(3)_c\otimes SU(2)_L\otimes U(1)_Y\equiv G_{SM}$, with $SU(2)_L\otimes U(1)_Y$ hidden\cite{sm} and $SU(3)_c$ confined\cite{su3}, it fails to explain several issues
like hierarchical fermion masses and mixing angles, charge quantization, CP
violation, replication of families, etc.. For example in the weak basis, before the symmetry is broken, the families in the SM are identical to each other; when the symmetry breaking takes place, the fermions acquire masses according to their experimental values and the three families become different. However in the SM there is not a mechanism for explaining the origin of families neither the fermion mass spectrum.

These drawbacks of the SM have led to a strong belief that the model is
still incomplete and that it must be regarded as a low-energy effective
field theory originating from a more fundamental one. That belief lies on
strong conceptual indications for physics beyond the SM which have
produced a variety of theoretically well motivated extensions of the
model: left-right symmetry, grand unification, supersymmetry, superstring
inspired extensions, etc.\cite{ext}.

At present, however, there are not definitive experimental facts that
point toward what lies beyond the SM and the best approach may be to
depart from it as little as possible. In this regard, $SU(3)_L\otimes
U(1)_X$ as a flavor group has been considered several times in the
literature; first as a family independent theory \cite{lee}, and then as a
family structure \cite{fp,long} which, among its best features, provides
with an alternative to the problem of the number $N$ of families, in the
sense that anomaly cancelation is achieved when $N$ is a multiple of
three; further, from the condition of $SU(3)_c$ asymptotic freedom which 
is valid only if the number of families is less than five, it follows that 
in those models $N$ is equal to 3.

Over the last decade two three family models based on the $SU(3)_c\otimes
SU(3)_L\otimes U(1)_X$ local gauge group have received special attention.
In one of them the three known left-handed lepton components for each
family are associated to three $SU(3)_L$ triplets as $(\nu_l, l^-, l^+)_L$,
where $l^+_L$ is related to the right-handed isospin singlet of the 
charged lepton $l_L^-$ in the SM \cite{fp}. In the other model the three 
$SU(3)_L$ lepton triplets are of the form $(\nu_l,l^-,\nu_l^c)_L$ where 
$\nu_{lL}^c$ is related to the right-handed component of the neutrino 
field $\nu_{lL}$\cite{long}. In the first model anomaly cancelation implies quarks with exotic electric charges $-4/3$ and 5/3, while in the second one the exotic particles have only ordinary electric charges.

A recent analysis of the $SU(3)_c\otimes SU(3)_L\otimes U(1)_X$ local
gauge theory (hereafter the 331 group) has shown that, by restricting the
fermion field representations to particles without exotic electric charges
and by paying due attention to anomaly cancelation, six different models
are obtained; while by relaxing the condition of nonexistence of exotic
electric charges, an infinite number of models can be generated\cite{ponce}.
Four of the six models without exotic electric charges are family models 
for quarks and leptons in which at least one of the 3 families is treated differently. The remaining two models are one family or family independent models. One of them, which was analyzed in Ref.\cite{spm}, is an $E_6$ subgroup. The other one, studied in this paper, is a subgroup of $SU(6)\otimes U(1)_X$, a new electroweak-strong unification group. For this last model we will do some phenomenological calculations in order to set the different mass scales and calculate the masses for all the spin 1/2 particles in one family.

This paper is organized as follows. In the next section we introduce
the model as an anomaly free theory based on the local gauge group
$SU(3)_c\otimes SU(3)_L\otimes U(1)_X$ and show that it is a subgroup of
an electroweak-strong unification group based on the gauge structure
$SU(6)\otimes U(1)_X$. In section three we describe the scalar sector
needed to break the symmetry and to produce masses to the fermion fields
in the model. In section four we study the gauge boson sector paying
special attention to the neutral currents present in the model and their
mixing. In section five we analyze the fermion mass spectrum. In section
six we use experimental results in order to constraint the mixing angle
between the two neutral currents and the mass scale of the new neutral
gauge boson. In the last section we summarize the model, do a comparison 
between the model in Ref.\cite{spm} and the present one and state our
conclusions. A technical appendix on the diagonalization of the $4\times 4$ mass matrix for the charged leptons in the model is presented at the end.

\section{The fermion content of the model} 
\subsection{$SU(3)_c\otimes SU(3)_L\otimes U(1)_X$ as a one-family anomaly 
free model}

In what follows we assume that the electroweak gauge group is
$SU(3)_L\otimes U(1)_X$ which contain $SU(2)_L\otimes U(1)_Y$ as a
subgroup. We also assume that the left handed quarks (color triplets) and
left-handed leptons (color singlets) transform as the $\bar{3}$ and $3$
representations of $SU(3)_L$ respectively. We have an $SU(3)_c$ 
vectorlike as in the SM and, contrary to most of the 331 models in the literature \cite{fp,long}, in our model the anomalies cancel individually in each family as it is done for the SM.

The most general expression for the electric charge generator in
$SU(3)_L\otimes U(1)_X$ is a linear combination of the three diagonal
generators of the gauge group 
\begin{equation}\label{ch}
Q=aT_{3L}+\frac{2}{\sqrt{3}}bT_{8L}+XI_3, 
\end{equation} 
where $T_{iL}=\lambda_{iL}/2, \,\lambda_{iL}$ being the Gell-Mann matrices
for $SU(3)_L$ normalized as {tr}$(\lambda_{iL}\lambda_{jL})=2\delta_{ij}$, 
and $I_3=Dg(1,1,1)$ is the diagonal $3\times 3$ unit
matrix. Choosing $a=1$ gives the usual isospin of the electroweak
interactions. $X$ and $b$ are arbitrary parameters to be fixed next.

We start by defining two $SU(3)_L$ triplets
\[\chi_L=\left(\begin{array}{c}d\\u\\q \end{array}\right)_L , \hspace{1cm}
\psi_L=\left(\begin{array}{c}\nu_e\\e^- \\l\end{array}\right)_L ,\] 
where $q_L$ and $l_L$ are $SU(2)_L$ singlet exotic quark and lepton fields
respectively. Now, if the ($SU(3)_L, U(1)_X$) quantum numbers for $\chi_L$ 
and $\psi_L$ are ($\bar{3},X_\chi$) and ($3,X_\psi$) respectively, 
then by using eq. (\ref{ch}) we have the relationship 
\begin{equation}\label{chql}
X_{\chi}+X_\psi=Q_q+Q_l=-1/3, 
\end{equation}
where $Q_q$ and $Q_l$ are the 
electric charges of the $SU(2)_L$ singlets $q_L$ and $l_L$ respectively, 
in units of the absolute value of the electron electric charge. 

Now, in order to cancel the $[SU(3)_L]^3$ anomaly two more $SU(3)_L$
lepton triplets with quantum numbers \{$3,X_i\}\;, i=1,2$, must
be introduced (together with their corresponding right-handed charged
components). Each one of those multiplets must include one $SU(2)_L$
doublet and one singlet of new exotic leptons.
The quarks fields $u^c_L,\; d^c_L$ and $q^c_L$ color anti-triplets and 
$SU(3)_L$ singlets, with $U(1)_X$ quantum numbers $X_u,\; X_d$ and $X_q$ 
respectively, must also be introduced in order to cancel the $[SU(3)_c]^3$ 
anomaly. Then the hypercharges $X_\alpha$ with $\alpha=\chi ,\psi
,1,2,u,d,q,...$ are fixed using Eqs. (\ref{ch}), (\ref{chql}) and the
anomaly constraint equations coming from the vertices 
$[SU(3)_C]^2U(1)_X$, $[SU(3)_L]^2U(1)_X$, $[grav]^2U(1)_X$ and
$[U(1)_X]^3$, where $[grav]^2U(1)_X$ stands for the gravitational 
anomaly \cite{del}. These equations are:
\begin{eqnarray} \nonumber
[SU(3)_c]^2U(1)_X &:& 3X_\chi+X_u+X_d+X_q=0\\ \nonumber
[SU(3)_L]^2U(1)_X &:& 3X_\chi+X_\psi+X_1+X_2=0\\ \nonumber
[grav]^2U(1)_X  &:& 9X_\chi+3X_u+3X_d+3X_q+3X_\psi+3X_1+3X_2+ 
                  \sum_{singl}X_{ls}=0\\  \nonumber
[U(1)_X]^3      &:&
9X_\chi^3+3X_u^3+3X_d^3+3X_q^3+3X_\psi^3+3X_1^3+3X_2^3+\sum_{singl}
X^3_{ls}=0, \end{eqnarray}
where $X_{ls}$ are the hypercharges of the right-handed charged lepton
singlets needed in order to have a consistent field theory.

In order to fix the parameter $b$, let us use for $q_L$ an exotic up type
quark $U$ of electric charge $Q_q=Q_U=2/3$. This fixes $b=1/2$, which
implies $Q_l=-1$ and then $l_L$ will be an exotic electron $E^-_L$. Then 
we have

\[\chi_L=\left(\begin{array}{c}d\\u\\U \end{array}\right)_L, \hspace{1cm}
\psi_L=\left(\begin{array}{c}\nu_e\\ e^-\\E_1^-\end{array}\right)_L. \]
\noindent 

In the former analysis we have used in the symmetry breaking chain
$SU(3)_L\longrightarrow SU(2)_L
\otimes U(1)_Z$ the branching rule $3\longrightarrow 2(1/6)+1(-1/3)$,
where the numbers in parenthesis are the new $Z$ hypercharge values. 
Then, by using the electric charge generator in Eq.(\ref{ch}) that we 
now may write as $Q=T_{3L}+Z+X$, we have that
$X_\chi=1/3$, $X_u=X_U=-2/3$ and $X_d=1/3$. For those values the 
anomaly $[SU(3)_c]^2U(1)_X$ is automatically canceled. Using the 
same argument we obtain $X_\psi=-2/3$.  Then, the anomaly constraint 
equations imply:  
$X_1+X_2=-1/3$, $\sum_{singl}X^3_{ls}+3 X_1^3+3 X_2^3=20/9$ 
and $\sum_{singl}X_{ls}=3$. By demanding for leptons of electric 
charges $\pm 1$ and 0 only, we must introduce, for the simplest 
solution, three equivalent singlets with hypercharges  
$X_{ls}=1$, $l=1,2,3$. Then $X_1=1/3$ and $X_2=-2/3$. 

We thus end up with the following anomaly free multiplet structure:
\[\begin{array}{||c|c|c|c||}\hline\hline
\chi_L=\left(\begin{array}{c}d\\u\\U \end{array}\right)_L & d^c_L & u^c_L&
U^c_L \\ \hline (3,\bar{3},{1\over 3}) & 
(\bar{3},1,{1\over 3}) & (\bar{3},1,-{2\over 3})
& (\bar{3},1,-{2\over 3}) \\ \hline\hline \end{array} \]

\[\begin{array}{||c|c|c|c|c|c||}\hline\hline
\psi_L=\left(\begin{array}{c} \nu_e\\ e^-\\E^-_1\end{array}\right)_L & 
\psi_{1L}=\left(\begin{array}{c} E^+_2\\ N_1^0\\ \nu^c_e\end{array}
\right)_L & \psi_{2L}=\left(\begin{array}{c}
N_2^0\\E^-_2\\E^-_3\end{array}\right)_L & e^+_L & E_{1L}^+ & E_{3L}^+ \\ \hline
(1,3,-{2\over 3}) & (1,3,{1\over 3}) & (1,3,-{2\over 3})
& (1,1,1) & (1,1,1) & (1,1,1) \\ \hline\hline
\end{array}  \]
The numbers in parenthesis refer to the ($SU(3)_C, SU(3)_L, U(1)_X$) 
quantum numbers respectively.

\subsection{$SU(3)_C\otimes SU(3)_L\otimes U(1)_X$ as an 
$SU(6)\otimes U(1)_X$ subgroup}

In a model based on a local gauge group $SU(6)\otimes U(1)_{\alpha}$, the
minimal choice of chiral multiplets that are free of the $[SU(6)]^3$
anomaly is $\bar{6} \oplus \bar{6} \oplus 15$ \cite{georgi}. If we want to
apply this group structure to the model in section 2.1, three $SU(6)$
singlets with identical hypercharges $\alpha_1$ must be introduced. 
The anomaly constraint equations read now  
\begin{eqnarray} \nonumber
[SU(6)]^2U(1)_{\alpha} &:& \alpha_6 + \alpha'_6 + 4 \alpha_{15}=0\\
\nonumber [grav]^2U(1)_{\alpha} &:& 6 \alpha_6 + 6 \alpha'_6 + 15
\alpha_{15} + 3\alpha_1=0\\ \nonumber [U(1)_{\alpha}]^3 &:& 6 \alpha^3_6 +
6 \alpha'^3_6 + 15 \alpha^3_{15} + 3\alpha^3_1=0, \end{eqnarray} 
where $\alpha_6$, $\alpha'_6$ and $\alpha_{15}$ are the $U(1)_{\alpha}$
hypercharges for the representations $\bar{6}$, $\bar{6}$ and $15$
respectively. Since a priori there is no way to distinguish between the
two $\overline{6}$ representations, we assume that $\alpha_6 = \alpha'_6$
and so the solution to the former set of equations is: $\alpha_1 = 3
\alpha_{15} = -{3 \over 2}\alpha_6 \equiv \alpha$, where $\alpha$ is a
free parameter.

Using for the symmetry breaking chain $SU(6) \longrightarrow
SU(3)_c\otimes SU(3)_L\otimes U(1)_{\beta}$ the branching rules $\bar{6}
\longrightarrow (\bar{3},1)(-\beta)+(1,3)(\beta)$ and $15 \longrightarrow
(\bar{3},1)(2\beta)+(1,3)(-2\beta)+(3, \bar{3})(0)$ \cite{slansky}, we see
immediately that the model in section 2.1 for the gauge group
$SU(3)_C\otimes SU(3)_L\otimes U(1)_X$ is a subgroup of the gauge group
$SU(6)\otimes U(1)_{\alpha}$ as far as we identify $\alpha = X = 1$. The
particle content of the irreducible representations in $SU(6)$ are \\ 

\begin{tabular}{lcc}
$\bar 6$=$\left(\begin{array}{c} u^c_1\\u^c_2\\u^c_3\\ \nu_e\\e^-\\E^-_1 \end{array}\right)_L,$  & 
$\left(\begin{array}{c} U^c_1\\U^c_2\\U^c_3\\N^0_2\\E^-_2\\E^-_3 \end{array}\right)_L ;$ &  $15$=$\left(\begin{array}{cccccc} 0 & d^c_1 &
- d^c_2 & d_1 & u_1 & U_1 \\ -d^c_1 & 0 &  d^c_3 & d_2 & u_2 & U_2 \\ d^c_2 & -d^c_3 & 0 & d_3 & u_3 & U_3  \\ -d_1 & -d_2 &  -d_3 & 0 & \nu^c_e & E^+_2 \\ -u_1 & -u_2 &  -u_3 & -\nu^c_e & 0 & N^0_1 \\ -U_1 & -U_2 &  -U_3 & -E^+_2 & -N^0_1 & 0 \end{array} \right)_L$; \\ \\
$1$=\hspace{0.3cm} $e^+_L,$ & $E^+_{1L},$ & $E^+_{3L}.$ \\
\end{tabular}\\

Since our $SU(6)\otimes U(1)_X$ is not a subgroup of $E_6$, the model 
introduced here is not related to the superstring derived flipped $SU(6)$ model \cite{flipped}.

\section{The scalar sector} 

Our aim is to break the symmetry following the pattern 
\[SU(3)_c\otimes
SU(3)_L\otimes U(1)_X\longrightarrow SU(3)_c\otimes SU(2)_L\otimes
U(1)_Y\longrightarrow SU(3)_c\otimes U(1)_Q\] 
\noindent
and give, at the same time, masses to the fermion fields in the model. 
With this in mind we introduce the following two Higgs scalars: $\phi_1(1,3,1/3)$ with a Vacuum Expectation Value (VEV) aligned in the direction $\langle\phi_1\rangle=(0,0,V)^T$ and $\phi_2(1,3,1/3)$ with a VEV aligned as $\langle\phi_2\rangle=(0,v/\sqrt{2},0)^T$, with the hierarchy 
$V >v\sim 250$ GeV (the electroweak breaking scale). The scale of $V$ is going to be fixed phenomenologically in section 6.

Note that $\phi_1$ and $\phi_2$ have the same quantum numbers but they get VEVs at different mass scales. This is necessary in order to give a large mass to the exotic quark and a realistic mass to the known up quark via a mixing with the exotic quark, as we will see in section 5.

One more Higgs scalar $\phi_3(1,3,-2/3)$ is needed in order to give a mass 
to the down quark field in the model but, as we will discuss in section 5, the most convenient VEV for this new scalar is zero $(\langle\phi_3\rangle=0)$.

\section{The gauge boson sector}
In the model there are a total of 17 gauge bosons: One gauge field
$B^\mu$ associated with $U(1)_X$, the 8 gluon fields associated
with $SU(3)_c$ which remain massless after breaking the symmetry, and 
other 8 associated with $SU(3)_L$ which, for $b=1/2$ in Eq.(\ref{ch}), 
can be written as:
\[{1\over 2}\lambda_\alpha A^\mu_\alpha={1\over \sqrt{2}}\left(
\begin{array}{ccc}D^\mu_1 & W^{+\mu} & K^{+\mu} \\ W^{-\mu} &
D^\mu_2 &  K^{0\mu} \\
K^{-\mu} & \bar{K}^{0\mu} & D^\mu_3 \end{array}\right) \]
where $D^\mu_1=A_3^\mu/\sqrt{2}+A_8^\mu/\sqrt{6},\;
D^\mu_2=-A_3^\mu/\sqrt{2}+A_8^\mu/\sqrt{6}$,
and $D^\mu_3=-2A_8^\mu/\sqrt{6}$. $\lambda_i, \; i=1,2,...,8$, are the eight Gell-Mann matrices normalized as mentioned in section 2.1. 

After breaking the symmetry with 
$\langle\phi_1\rangle + \langle\phi_2\rangle$  and using for the covariant
derivative for triplets
$D^\mu=\partial^\mu-i{g\over 2} \lambda_\alpha
A^\mu_\alpha-ig'XB^\mu$, we get the following mass terms for the
charged gauge bosons:
$M^2_{W^\pm}={g^2\over 4}v^2, \; M^2_{K^\pm}={g^2\over 2}V^2$, and
$M^2_{K^0(\bar{K}^0)}={g^2\over 2}(V^2 + v^2/2)$.
For the three neutral gauge bosons we get  mass terms of the form:
\[M=V^2(\frac{g'B^\mu}{3}-\frac{gA_8^\mu}{\sqrt{3}})^2 +
\frac{v^2}{8}(\frac{2g'B^\mu}{3}-gA^\mu_3
+\frac{gA_8^\mu}{\sqrt{3}})^2.\]
By diagonalizing $M$ we get the physical neutral gauge bosons which are
defined through the mixing angle $\theta$ and $Z_\mu,\; Z'_\mu$ by:
\begin{eqnarray}\nonumber
Z_1^\mu&=&Z_\mu \cos\theta+Z'_\mu \sin\theta \; ,\\ \nonumber
Z_2^\mu&=&-Z_\mu \sin\theta+Z'_\mu \cos\theta,
\end{eqnarray}
where
\begin{equation} \label{tan}
\tan(2\theta) =- \frac{v^2(1-2 S_W^2)\sqrt{3-4 S_W^2}}
{4 C_W^4 V^2- v^2(1-2 S^4_W)}.
\end{equation}
The photon field $A^\mu$ and the fields $Z_\mu$ and $Z'_\mu$ are given by
\begin{eqnarray} \nonumber
A^\mu&=&S_W A_3^\mu + C_W\left[\frac{T_W}{\sqrt{3}}A_8^\mu+
(1-T_W^2/3)^{1/2}B^\mu\right]\; , \nonumber \\  Z^\mu&=& C_W
A_3^\mu - S_W\left[\frac{T_W}{\sqrt{3}}A_8^\mu+
(1-T_W^2/3)^{1/2}B^\mu\right] \; , \nonumber \\ \label{fzzp}
Z'^\mu&=&-(1-T_W^2/3)^{1/2}A_8^\mu+\frac{T_W}{\sqrt{3}}B^\mu  \; .
\end{eqnarray}
$S_W=\sqrt{3}g'/\sqrt{3g^2+4g'^2}$ and $C_W$ are the sine and
cosine of the electroweak mixing angle respectively and
$T_W=S_W/C_W$. We can also identify the $Y$ hypercharge associated
with the SM gauge boson as:
\begin{equation}\label{y}
Y^\mu=\left[\frac{T_W}{\sqrt{3}}A_8^\mu+
(1-T_W^2/3)^{1/2}B^\mu\right].
\end{equation}

\subsection{Charged currents}
The Hamiltonian for the charged currents can be written as 
\begin{eqnarray}\nonumber
H^{CC}&=&{g\over \sqrt{2}}[W^+_\mu(-\bar{u}_L\gamma^\mu d_L+
\bar{\nu}_{eL}\gamma^\mu e^-_L+\bar{N}^0_{2L}\gamma^\mu E^-_{2L}+
\bar{E}^+_{2L}\gamma^\mu N^0_{1L}) \\ \nonumber & &
+K^+_\mu(-\bar{U}_L\gamma^\mu d_L+ \bar{\nu}_{eL}\gamma^\mu
E^-_{1L}+\bar{N}^0_{2L}\gamma^\mu E^-_{3L}+ \bar{E}^+_{2L}\gamma^\mu
\nu^c_{eL}) \\ & & +K^0_\mu(-\bar{U}_L\gamma^\mu u_L+
\bar{N}^0_{1L}\gamma^\mu \nu^c_{eL}+\bar{E}^-_{2L}\gamma^\mu
E^-_{3L}+ \bar{e}^-_L\gamma^\mu E^-_{1L})] + H.c.,
\end{eqnarray}
which implies that the interactions with the $K^\pm$ and
$K^0(\bar{K}^0)$ bosons violate the lepton number and the weak
isospin. Notice also that the first two terms in the previous
expression constitute the charged weak current of the SM as far as
we identify $W^\pm$ as the $SU(2)_L$ charged left-handed weak
bosons.

\subsection{Neutral currents}
The neutral currents $J_\mu(EM),\; J_\mu(Z)$ and $J_\mu(Z')$
associated with the Hamiltonian $H^0=eA^\mu J_\mu(EM)+{g\over
{C_W}}Z^\mu J_\mu(Z) + {g'\over \sqrt{3}}Z'^\mu J_\mu(Z')$ are:
\begin{eqnarray}\nonumber
J_\mu(EM)&=&{2\over 3}\bar{u}\gamma_\mu u-{1\over
3}\bar{d}\gamma_\mu d +{2\over 3}\bar{U}\gamma_\mu U-
\bar{e}^-\gamma_\mu e^-- \bar{E}^-_1\gamma_\mu E^-_1- \bar{E}^-_2\gamma_\mu E^-_2
- \bar{E}^-_3\gamma_\mu E^-_3   \\ \nonumber
&=&\sum_f q_f\bar{f}\gamma_\mu f \\ \nonumber
J_\mu(Z)&=&J_{\mu,L}(Z)-S^2_WJ_\mu(EM)\\
J_\mu(Z')&=&T_WJ_\mu(EM)-J_{\mu,L}(Z'),
\end{eqnarray}
\noindent where $e=gS_W=g'C_W\sqrt{1-T_W ^2/3}>0$ is the electric charge, $q_f$
is the electric charge of the fermion $f$ in units of $e$, $J_\mu(EM)$
is the electromagnetic current, and the left-handed currents are
\begin{eqnarray} \nonumber
J_{\mu,L}(Z)&=&{1\over 2}(\bar{u}_L\gamma_\mu u_L-
\bar{d}_L\gamma_\mu d_L+\bar{\nu}_{eL}\gamma_\mu \nu_{eL}-
\bar{e}^-_L\gamma_\mu e^-_L+\bar{N}^0_2\gamma_\mu N^0_2
-\bar{E}^-_2\gamma_\mu E^-_2) \\ \nonumber &=&\sum_f
T_{3f}\bar{f}_L\gamma_\mu f_L ,\\ \nonumber
J_{\mu,L}(Z')&=&S_{2W}^{-1}(-\bar{d}_L\gamma_\mu d_L +
\bar{\nu}_{eL}\gamma_\mu \nu_{eL}
+\bar{E}^+_{2L}\gamma_\mu E^+_{2L}
+\bar{N}^0_{2L}\gamma_\mu N^0_{2L})
\\ \nonumber &+&
T_{2W}^{-1}(-\bar{u}_L\gamma_\mu u_L+\bar{e}^-_L\gamma_\mu e^-_L
+\bar{N}^0_{1L}\gamma_\mu N^0_{1L}
+\bar{E}^-_{2L}\gamma_\mu E^-_{2L}) 
\\ \nonumber &-& T_W^{-1}(-\bar{U}_L\gamma_\mu U_L +
\bar{E}^-_{1L}\gamma_\mu E^-_{1L} +\bar{\nu^c}_{eL}\gamma_\mu \nu^c_{eL}
+\bar{E}^-_{3L}\gamma_\mu E^-_{3L}) \\ &=&\sum_f
T_{9f}\bar{f}_L\gamma_\mu f_L,
\end{eqnarray}
where $S_{2W}=2S_WC_W,\; T_{2W}=S_{2W}/C_{2W}, \; C_{2W}=C^2_W-S^2_W, \;
\bar{N}^0_{2}\gamma_\mu N^0_{2}=\bar{N}^0_{2L}\gamma_\mu N^0_{2L}
+\bar{N}^0_{2R}\gamma_\mu N^0_{2R}= \bar{N}^0_{2L}\gamma_\mu
N^0_{2L} -\bar{N}^{0c}_{2L}\gamma_\mu N^{0c}_{2L}=
\bar{N}^0_{2L}\gamma_\mu N^0_{2L} -\bar{N}^0_{1L}\gamma_\mu
N^0_{1L}$, and similarly $\bar{E}_2\gamma_\mu E_2=\bar{E}^-_{2L}\gamma_\mu
E^-_{2L} - \bar{E}^+_{2L}\gamma_\mu E^+_{2L}$. In this way
$T_{3f}=Dg(1/2,-1/2,0)$ is the third component of the weak isospin
and $T_{9f}=Dg(S_{2W}^{-1}, 
T_{2W}^{-1},-T_W^{-1})$ is a convenient $3\times 3$ diagonal 
matrix, acting both of them on the representation 3 of $SU(3)_L$. 
Notice that $J_\mu(Z)$ is just the generalization of the neutral current 
present in the SM. This allows us to identify $Z_\mu$ as the neutral 
gauge boson of the SM, which is consistent with 
Eqs.(\ref{fzzp}) and (\ref{y}).

The couplings of the mass eigenstates $Z_1^\mu$ and $Z_2^\mu$ are given by:
\begin{eqnarray} \nonumber
H^{NC}&=&\frac{g}{2C_W}\sum_{i=1}^2Z_i^\mu\sum_f\{\bar{f}\gamma_\mu
[a_{iL}(f)(1-\gamma_5)+a_{iR}(f)(1+\gamma_5)]f\} \\ \nonumber
      &=&\frac{g}{2C_W}\sum_{i=1}^2Z_i^\mu\sum_f\{\bar{f}\gamma_\mu
      [g(f)_{iV}-g(f)_{iA}\gamma_5]f\},
\end{eqnarray}
where
\begin{eqnarray} \nonumber
a_{1L}(f)&=&\cos\theta(T_{3f}-q_fS^2_W)-\frac{g'\sin\theta
C_W}{g\sqrt{3}} (T_{9f}-q_fT_W)\;, \\ \nonumber
a_{1R}(f)&=&-q_fS_W(\cos\theta
S_W-\frac{g'\sin\theta}{g\sqrt{3}})\;,\\ \nonumber
a_{2L}(f)&=&-\sin\theta(T_{3f}-q_fS^2_W)-\frac{g'\cos\theta
C_W}{g\sqrt{3}} (T_{9f}-q_fT_W)\;, \\ \label{a}
a_{2R}(f)&=&q_fS_W(\sin\theta S_W+\frac{g'\cos\theta}{g\sqrt{3}})
\end{eqnarray}
and
\begin{eqnarray} \nonumber
g(f)_{1V}&=&\cos\theta(T_{3f}-2S_W^2q_f)-\frac{g'\sin\theta}{g\sqrt{3}}
(T_{9f}C_W-2q_fS_W)\;, \\ \nonumber
g(f)_{2V}&=&-\sin\theta(T_{3f}-2S_W^2q_f)-\frac{g'\cos\theta}{g\sqrt{3}}
(T_{9f}C_W-2q_fS_W) \;,\\ \nonumber g(f)_{1A}&=&\cos\theta
T_{3f}-\frac{g'\sin\theta}{g\sqrt{3}}T_{9f}C_W\;, \\ \label{g}
g(f)_{2A}&=&-\sin\theta
T_{3f}-\frac{g'\cos\theta}{g\sqrt{3}}T_{9f}C_W.
\end{eqnarray}
The values of $g_{iV},\; g_{iA}$ with $i=1,2$ are listed in Tables I and II. 
As we can see, in the limit $\theta=0$ the couplings of $Z_1^\mu$ to 
the ordinary leptons and quarks are the same as in the SM, due to 
this we can test the new physics beyond the SM predicted by this 
particular model.

\vspace{1.5cm}

\begin{center}
TABLE I. The $Z_1^\mu\longrightarrow \bar{f}f$ couplings.
\begin{tabular}{||l||c|c||}\hline\hline
f& $g(f)_{1V}$ & $g(f)_{1A}$ \\ \hline\hline d& $(-{1\over
2}+{2S_W^2\over 3})[\cos\theta-\sin\theta /(4C_W^2-1)^{1/2}]$ &
$-{1\over
2}(\cos\theta-\sin\theta/[2(4C_w^2-1)^{1/2}])$ \\ \hline u&
$\cos\theta ({1\over 2}-{4S_W^2 \over 3})+
\frac{\sin\theta}{(4C_W^2-1)^{1/2}}({1\over 2} + {S_W^2\over 3})$
& ${1\over 2} \cos\theta + \sin\theta
(1/2 - S_W^2)/(4C_W^2-1)^{1/2}$ \\ \hline U& $-{4S_W^2\cos\theta
\over 3}-\sin\theta (1-{7\over 3}S_W^2)/(4C_W^2-1)^{1/2}$ &
$-C_W^2\sin\theta /(4C_W^2-1)^{1/2}$ \\ \hline $e^-$& $\cos\theta
(-{1\over 2}+2S_W^2)- \frac{\sin\theta}{(4C_W^2-1)^{1/2}}({1\over
2} + S_W^2)$ & $ -{\cos\theta\over
2}-\frac{\sin\theta}{(4C_W^2-1)^{1/2}}({1\over 2}-S_W^2)$
 \\ \hline
$E^-_1 ,E^-_3$& $2\cos\theta S_W^2 +
\frac{\sin\theta}{(4C_W^2-1)^{1/2}}(1 - 3S_W^2)$ &
$C_W^2\sin\theta /(4C_W^2-1)^{1/2}$ \\
\hline
$E^-_2$& $\cos\theta (-1+2S_W^2)-
\frac{S_W^2\sin\theta}{(4C_W^2-1)^{1/2}}$ &
$-C_W^2\sin\theta /(4C_W^2-1)^{1/2}$ \\ \hline
$\nu_e, N_2^0$ &
${1\over 2}[\cos\theta-\sin\theta/(4C_W^2-1)^{1/2}]$ &
$ {1\over 2}[\cos\theta-\sin\theta/(4C_W^2-1)^{1/2}]$ \\ \hline
$\nu_e^c$ & $C_W^2\sin\theta /(4C_W^2-1)^{1/2}$ &
$C_W^2\sin\theta /(4C_W^2-1)^{1/2}$ \\ \hline
$N_1^0$ &
$-{1\over 2}\cos\theta-{\sin\theta \over (4C_W^2-1)^{1/2}}
(1/2 - S_W^2)$& $-{1\over 2}\cos\theta-{\sin\theta \over 
(4C_W^2-1)^{1/2}}(1/2 - S_W^2)$
\\ \hline\hline
\end{tabular}
\end{center}


\vspace{1.5cm}

\begin{center}
TABLE II. The $Z_2^\mu\longrightarrow \bar{f}f$ couplings.
\begin{tabular}{||l||c|c||}\hline\hline
f& $g(f)_{2V}$ & $g(f)_{2A}$ \\ \hline\hline d& $({1\over
2}-{2S_W^2\over 3})[\sin\theta+\cos\theta /(4C_W^2-1)^{1/2}]$ &
${1\over
2}(\sin\theta+\cos\theta/[2(4C_w^2-1)^{1/2}])$ \\ \hline u&
$-\sin\theta ({1\over 2}-{4S_W^2 \over 3})+
\frac{\cos\theta}{(4C_W^2-1)^{1/2}}({1\over 2} + {S_W^2\over 3})$
& $-{1\over 2} \sin\theta + \cos\theta
(1/2 - S_W^2)/(4C_W^2-1)^{1/2}$ \\ \hline U& ${4S_W^2\sin\theta
\over 3}-\cos\theta (1-{7\over 3}S_W^2)/(4C_W^2-1)^{1/2}$ &
$-C_W^2\cos\theta /(4C_W^2-1)^{1/2}$ \\ \hline $e^-$& $\sin\theta
(-{1\over 2}+2S_W^2)- \frac{\cos\theta}{(4C_W^2-1)^{1/2}}({1\over
2} + S_W^2)$ & $ {\sin\theta\over
2}-\frac{\cos\theta}{(4C_W^2-1)^{1/2}}({1\over 2}-S_W^2)$
 \\ \hline
$E^-_1 ,E^-_3$& $-2\sin\theta S_W^2 +
\frac{\cos\theta}{(4C_W^2-1)^{1/2}}(1 - 3S_W^2)$ &
$C_W^2\cos\theta /(4C_W^2-1)^{1/2}$ \\
\hline
$E^-_2$& $-\sin\theta (-1+2S_W^2)-
\frac{S_W^2\cos\theta}{(4C_W^2-1)^{1/2}}$ &
$-C_W^2\cos\theta /(4C_W^2-1)^{1/2}$ \\ \hline
$\nu_e, N_2^0$ & $-{1\over 2}[\sin\theta+\cos\theta/(4C_W^2-1)^{1/2}]$ &
$-{1\over 2}[\sin\theta+\cos\theta/(4C_W^2-1)^{1/2}]$ \\ \hline
$\nu_e^c$ &
$C_W^2\cos\theta /(4C_W^2-1)^{1/2}$  &
$C_W^2\cos\theta /(4C_W^2-1)^{1/2}$  \\ \hline
$N_1^0$ &
${1\over 2}\sin\theta-{\cos\theta \over (4C_W^2-1)^{1/2}}
(1/2 - S_W^2)$& ${1\over 2}\sin\theta-{\cos\theta \over 
(4C_W^2-1)^{1/2}}(1/2 - S_W^2)$
\\ \hline\hline
\end{tabular}
\end{center}
\vspace{1.5cm}


\section{Fermion masses}
The Higgs scalars introduced in section 3 not only break the symmetry in an
appropriate way, but produce the following mass terms for the fermions of
the model:
\subsection{Masses for the up quark sector}
For the quark sector we can write the following Yukawa terms:
\begin{equation}\label{yuq}
{\cal L}_Y^Q=\chi_L^TC(h_u\phi_2u_L^c+h_{U}\phi_1U_L^c
+h_{uU}\phi_2U_L^c+h_{Uu}\phi_1u_L^c) + H.c.,
\end{equation}
where $h_\eta, \; \eta = u, U, uU, Uu$, are Yukawa
couplings of order one and $C$ is the charge conjugate operator. From 
Eq.(\ref{yuq}) we get for the up quark sector a mass matrix
in the basis $(u,U)_L$ of the form:
\begin{equation}\label{qm}
M_{uU}=\left(\begin{array}{cc}
h_{u}v/\sqrt{2} & h_{uU}v/\sqrt{2} \\
h_{Uu}V & h_{U}V \end{array} \right).
\end{equation}
For the particular case $h_{u}=h_{uU}=h_{Uu}=h_{U}\equiv h$, the
mass eigenvalues of the previous matrix are $m_u=0$ and
$m_U=h(V+v/\sqrt{2})$. Since there is not a physical reason for
the Yukawas to be equal, let us calculate the mass eigenvalues as
a function of $|M_{uU}|\equiv (h_uh_U-h_{uU}h_{Uu})$ the
determinant of the Yukawas, and in the expansion $v/V$. The
algebra shows that $m_U\simeq
h_{U}V+h_uv/\sqrt{2}-|M_{uU}|v^2/\sqrt{2}Vh_{U}$ and
$m_u\simeq
v|M_{uU}|(1+h_{uU}h_{Uu}v/\sqrt{2}Vh_{U}^3)/\sqrt{2}h_U$, so a large mass is generated for the exotic up quark and for the other one, associated with the ordinary up quark, a mass lowered by the determinant $|M_{uU}|$ which we expect to be smaller than one.

\subsection{Lepton masses}
For the lepton sector we have the following Yukawa terms:
\begin{eqnarray}\nonumber
{\cal L}_Y^l&=&\epsilon_{abc}[\psi_L^aC(h_1\psi_{1L}^b\phi_1^c+h_2\psi_{1L}^b
\phi_2^c)
+\psi_{1L}^aC(h_3\psi_{2L}^b\phi_1^c+h_4\psi_{2L}^b\phi_2^c)]\\ 
\nonumber & &
+ \psi_L^TC \phi_1^*(h_5 e^+_L +h_6 E^+_{1L}+h_7 E^+_{3L})
+ \psi_{2L}^TC \phi_1^*(h_8 e^+_L +h_9 E^+_{1L}+h_{10} E^+_{3L})\\ \nonumber & &
+\psi_L^TC \phi_2^*(h_{11} e^+_L +h_{12} E^+_{1L}+h_{13} E^+_{3L})
+\psi_{2L}^TC \phi_2^*(h_{14} e^+_L +h_{15} E^+_{1L}+h_{16} E^+_{3L}) \\ 
& &
+H.c., \label{malep}
\end{eqnarray}

\noindent where $a,b,c$ are $SU(3)_L$ tensor indices and the Yukawas are
again of order one. 

\subsubsection{Masses for the neutral leptons}
For the neutral leptons Eq.(\ref{malep}) produce, in the basis
$(\nu_e,\nu^c_e,N_1,N_2)$, the mass matrix

\begin{equation}\label{nm}
M_N=\left(\begin{array}{cccc}
0 & -h_2v/\sqrt{2} &  h_1V & 0\\
-h_2v/\sqrt{2} & 0 & 0 & h_4v/\sqrt{2} \\
h_1V & 0 & 0 & -h_3V \\
 0 & h_4v/\sqrt{2} & -h_3V & 0 
\end{array}\right).
\end{equation}

The eigenvalues for this matrix are  

\[m_1 =\pm \sqrt{{1 \over 2}(A+\sqrt{A^2-4D^2})}, \hspace{1.0cm} m_2 = \pm
\sqrt{{1 \over 2}(A-\sqrt{A^2-4D^2})},\] 
\noindent where $A=V^2(h^2_1+h^2_3)+(v^2/2)(h^2_2+h^2_4)$ and
$D^2=V^2v^2(h_1h_4-h_2h_3)/2$. These values imply for this model that
there are two Dirac neutrinos for each family. For the particular case
$D=0$ we find $m_1 =\pm V \sqrt{(h^2_1+h^2_3)+{v^2 \over {2V^2}}(h^2_2+h^2_4)}$
and $m_2 = 0$, which means a massless Dirac neutrino and, in the limit $V
\gg v$, a very massive one. In the case $D \neq 0$ and $\vert D \vert \ll
\vert A \vert$, an expansion of $m_i, i=1,2,$ in powers of $D^2/A^2$
gives: $m_1= \pm \sqrt{A}(1-D^2/2A^2+...) \approx \pm V
\sqrt{h^2_1+h^2_3}$ and $m_2= \pm D/\sqrt{A}(1+D^2/2A^2+...) \approx \pm v
(h_1h_4-h_2h_3)/ \sqrt{2(h^2_1+h^2_3)}$ (in the limit $V\gg v$), which
means that the mass of the light neutrino is suppressed with respect to
the scale $v$ by the small value of $h_1h_4-h_2h_3 \sim D$. Notice that $D
= 0$ if we impose in our model the symmetry $\phi_1 \longleftrightarrow
\pm \phi_2$ which implies $h_1= \pm h_2$ and $h_3= \pm h_4$.\\

\subsubsection{Masses for the charged leptons}
For the charged leptons Eq.(\ref{malep}) produces, in the basis 
$(e,E_1,E_2,E_3)$, the mass matrix 

\begin{equation}\label{cmb}
M_{eE}=\left(\begin{array}{cccc} h_{11}v/\sqrt{2} & h_{12}v/\sqrt{2} &
- h_{1}V & h_{13}v/\sqrt{2}\\ h_{5}V & h_{6}V &  h_{2}v/\sqrt{2} & h_{7}V\\ h_{14}v/\sqrt{2} & h_{15}v/\sqrt{2} &
- h_{3}V & h_{16}v/\sqrt{2}\\ h_8 V & h_9 V &  h_{4}v/\sqrt{2} & h_{10}V \end{array} \right),
\end{equation}

\noindent and we must find the eigenvalues for the matrix $M_{eE}M^T_{eE} \equiv M^2(v,V)$. 

Let us first consider the particular case $M^2(v=0,V)$ (see Appendix A). In this case we have first that $\det M^2(v=0,V)=0$, so, there is a zero 
mass eigenstate that we may identify as the electron (for the first family 
for example). Second, $M^2$ has three eigenvalues different from zero and of the order $V^2$, which means that at the first stage of the symmetry 
breaking chain 

\[SU(3)_c\otimes SU(3)_L\otimes U(1)_X\stackrel{\langle\phi_1\rangle}
{\longrightarrow} SU(3)_c\otimes SU(2)_L\otimes U(1)_Y,\] 

\noindent the three exotic charged leptons get heavy masses of order $V >> 
v\sim 250$ GeV (the electroweak breaking mass scale).

Now, for $v \ne 0$ and by taking profit of $v/V \ll 1$, we can diagonalize $M^2(v,V)$ using matrix perturbation theory up to first order in the perturbation \cite{schiff}. In Appendix A we show that, at this order, three eigenvalues of $M^2$ are of the order $V^2$ and the eigenvalue corresponding to the mass of the lightest charged lepton is given by 

\[m^2_e \simeq {h_1^2H^2 + h_3^2 J^2 - 2h_1h_3K^2 \over 2(h_1^2 + h_3^2)} v^2, \]
where $H^2=h_{14}^2 + h_{15}^2 + h_{16}^2$, $J^2=h_{11}^2 + h_{12}^2 + h_{13}^2$ and $K^2=h_{11}h_{14} + h_{12}h_{15} + h_{13}h_{16}$. Notice that with Yukawas of the order one $H^2 \approx J^2 \approx K^2$, and $m_e$ is suppressed with respect to the scale $v$ by a small difference of Yukawas, namely $h_1 - h_3$. It is worth noticing that the value of $m_e$ is independent of the symmetry $\phi_1 \longleftrightarrow \pm \phi_2$ which makes zero the mass of the light neutrino.

\subsection{Mass for the down quark}
The down quark does not get a mass at zero level. 
There are two alternatives if we want to produce a mass for this field:\\
1- Introducing a third Higgs scalar $\phi_3(1,3,-2/3)$ with a VEV 
$\langle\phi_3\rangle=(v'/\sqrt{2},0,0)^T$, where $v' \sim v$, will immediately produce a tree level mass term $m_d = h_d \; v'/\sqrt{2}$.\\
2- Introducing the same Higgs scalar $\phi_3(1,3,-2/3)$ but without a VEV 
($\langle\phi_3\rangle=0$ as mentioned in section 3) produces a 
one-loop radiative mass for the down quark $d$. Fig.1 shows one of the diagrams contributing to this mass.
A similar diagram is obtained by simultaneously replacing $\phi^+_1$ by $\phi^+_2$ and $\phi^0_2$ by $\phi^0_1$ in Fig.1.

In both cases $\phi_3$ does not introduce new mass terms for the leptons 
in the model. From the two alternatives discussed above, we prefer the second one since it is the only one generating a natural small mass value for the down quark $d$. For the first alternative we must explain either why $v'<< v$ or why $h_d$, the Yukawa coupling for the down quark, is much less than one.


 
\section{Constrains on the $(Z^\mu-Z'^{\mu})$ mixing angle and the
$Z^{\mu}_2$ mass}

To get bounds on $\sin\theta$ and $M_{Z_2}$ we use experimental 
parameters measured at the $Z$ pole from CERN $e^+e^-$ collider (LEP), SLAC 
Linear Collider (SLC) and data from atomic parity violation as given in 
table III \cite{lang2001,data}.
The expression for the partial decay width for $Z^{\mu}_1$ in two fermions is
\begin{equation}
\Gamma(Z^{\mu}_1\rightarrow f\bar{f})=\frac{N_C G_F
M_{Z_1}^3}{6\pi\sqrt{2}}\rho \left[\frac{3\beta-\beta^3}{2}
(g(f)_{1V})^2+\beta^3 (g(f)_{1A})^2\right](1+\delta_f)R_{QCD+QED},\label{ancho}
\end{equation}
\noindent 
where $f$ is an ordinary SM fermion with $m_f \leq M_{Z_1}/2$, $Z^{\mu}_1$ 
is the physical gauge boson observed at LEP, $N_C=3(1)$ is the number of 
colors of quarks (leptons), $R_{QCD+QED}$ are the QCD and QED corrections 
and $\beta=\sqrt{1-4 m_b^2/M_{Z_1}^2}$ is a kinematic factor. The factor 
$\delta_f$ contains the one loop vertex contributions and it 
is zero for all fermions except for the bottom quark for which it can be 
written as $\delta_b\approx 10^{-2} (-m_t^2/(2
M_{Z_1}^2)+1/5)$ \cite{pich}. The $\rho$ parameter has two
contributions, one of them is the oblique correction given by
$\delta\rho\approx 3G_F m_t^2/(8\pi^2\sqrt{2})$ and the other one is the
tree level contribution due to the $(Z_{\mu} - Z'_{\mu})$ mixing and can be 
parametrized as $\delta\rho_V\approx
(M_{Z_2}^2/M_{Z_1}^2-1)\sin^2\theta$. Finally, $g(f)_{1V}$ and
$g(f)_{1A}$ are the coupling constants of the $Z_1^{\mu}$ field with 
ordinary fermions and they are listed in table I. In the following we 
will use \cite{data}: $m_t=174.3$ GeV, $\alpha_s(m_Z)=0.1192$,
$\alpha(m_Z)^{-1}=127.938$, and $S^2_W=0.2333$.

The effective weak charge in atomic parity violation, $Q_W$, can be expressed as a function of the number of protons ($Z$) and the number of neutrons ($N$) in the atomic nucleus in the form
\begin{equation}
Q_W=-2\left[(2Z+N)c_{1u}+(Z+2N)c_{1d}\right],
\end{equation}
where $c_{1q}=2g(e)_{1A}g(q)_{1V}$. 
The theoretical value for $Q_W$ is given by \cite{marciano}
\begin{equation}
Q_W(^{133}_{55} C_s)=-73.09\pm 0.04 +\Delta Q_W.
\end{equation}
$\Delta Q_W$, which includes the contributions of new physics, 
can be written as \cite{durkin}
\begin{equation} \label{deltaq}
\Delta Q_W = \left[\left(1+4\frac{S_W^2}{1-2S_W^2}\right)-Z\right] 
\delta\rho_V +\Delta Q_W'.
\end{equation}
The term $\Delta Q_W'$ is model dependent and it can be obtained for
our model by using $g(e)_{1A}$ and $g(q)_{1V}$ from Table I. The value we obtain is
\begin{equation} 
\Delta Q_W'=(8.49 Z +5.59 N) \sin\theta - (3.10 Z + 2.62 N)
\frac{M^2_{Z_1}}{M^2_{Z_2}}\; .
\end{equation}

The discrepancy between the SM and the experimental data for $\Delta Q_W$ is given by \cite{todos}
\begin{eqnarray}
\Delta Q_W&=&Q_W^{exp}-Q_W^{SM}\nonumber \\ &=&1.03\pm 0.44\; .
\end{eqnarray}

Introducing the expressions for $Z$ pole observables in eq.(\ref{ancho}), 
with $\Delta Q_W$ in terms of new physics in eq.(\ref{deltaq}) and using experimental data from LEP, SLC and atomic parity violation (see table III),  we do a $\chi^2$ fit and we find the best allowed region for $\sin\theta$ vs. $M_{Z_2}$ at 95$\%$ CL.

\pagebreak
\begin{center} TABLE III. Experimental data and SM values for the parameters
\begin{tabular}{||l||l|l||} \hline\hline
& Experimental results & SM \\ \hline
$\Gamma_Z$ (GeV) & $2.4952\pm 0.0023$ & $2.4963\pm 0.0016$  \\ \hline
$\Gamma(had)$ (GeV) & $1.7444\pm 0.0020$ & $1.7427\pm 0.0015$  \\ \hline
$\Gamma(l^+l^-)$ (MeV) & $83.984\pm 0.086$ & $84.018\pm 0.028$ \\ \hline
$R_e$ & $20.804\pm 0.050$ & $20.743\pm 0.018$ \\ \hline
$A_{FB}(e)$ & $0.0145\pm 0.0025$ & $0.0165\pm 0.0003$  \\ \hline
$R_b$ & $0.21653\pm 0.00069$ & $0.21572\pm 0.00015$  \\ \hline 
$R_c$ & $0.1709\pm 0.0034$ & $0.1723\pm 0.0001$ \\ \hline
$A_{FB}(b)$ & $0.0990\pm 0.0020$ & $0.1039\pm 0.0009$ \\ \hline
$A_{FB}(c)$ & $0.0689\pm 0.0035$ & $0.0743\pm 0.0007$ \\ \hline
$A_{FB}(s)$ & $0.0976\pm 0.0114$ & $0.1040\pm 0.0009$ \\ \hline
$A_{b}$ & $0.922\pm 0.023$ & $0.9348\pm 0.0001$ \\ \hline
$A_{c}$ & $0.631\pm 0.026$ & $0.6683\pm 0.0005$ \\ \hline
$A_{s}$ & $0.82\pm 0.13$ & $0.9357\pm 0.0001$ \\ \hline
$A_{e}({\cal P}_{\tau})$ & $0.1498\pm 0.0048$ & $0.1483\pm 0.0012$
\\ \hline        
$Q_W(Cs)$ & $-72.06\pm 0.28\pm 0.34$  & $-73.09\pm 0.04$
\\ \hline\hline
\end{tabular} \end{center} \noindent

In Fig. 2 we display the allowed region for $\sin\theta$ vs.
$M_{Z_2}$ at 95$\%$ confidence level. The result is
\begin{eqnarray}
-0.00015\leq &\sin\theta& \leq 0  \;, \nonumber \\ 1.5\; TeV \leq &
M_{Z_2},
\end{eqnarray}
which shows that the mass of the new neutral gauge boson is compatible with
the bound got in $p\bar{p}$ collisions at the Tevatron \cite{fermi}. 
The negative value for $\theta$ obtained from phenomenology is consistent 
with the negative value on front of Eq.(\ref{tan}), which by the way implies 
$60\;TeV\leq V<\infty$, where the $\infty$ upper limit is clearly seen from  
Fig. 3 ($\sin\theta\rightarrow 0$ when $M_{Z_2}\rightarrow\infty)$.


\section{Concluding Remarks}
We have presented an anomaly-free model based on the local gauge
group $SU(3)_c\otimes SU(3)_L\otimes U(1)_X$ which is a subgroup
of an electroweak-strong unification group $SU(6)\otimes U(1)_X$, 
does not studied in the literature so far. We break the
gauge symmetry down to $SU(3)_c\otimes U(1)_{Q}$ and at the same
time give masses to the fermion fields in the model in a
consistent way by using three different Higgs scalars $\phi_i,\:
i=1,2,3$ with $\langle\phi_3\rangle =0$ and $\langle\phi_i\rangle\ne 0,\; 
i=1,2$ as stated in section 3. This Higgs fields and their VEVs 
set two different mass scales: $v\sim 250$ GeV $<<V$, where $V\geq 60$ TeV. 
By using experimental results from LEP, SLC and atomic
parity violation we bound the mixing angle and the mass of the
additional neutral current to be $-0.00015\leq\sin\theta<0$
and $1.5\; TeV\leq M_{Z_2}$  at 95$\%$ CL.

The mass spectrum
of the fermion fields in the model arises as a consequence
of the mixing between ordinary fermions and their exotic
counterparts without the need of different scale Yukawa couplings.
Conspicuously, the four neutral leptons in the model split as two
Dirac neutrinos one heavy and the other one light.

Notice that all the new states are vector-like with respect to the
SM quantum numbers. They consist of an isosinglet quark $U$ of
electric charge $2/3$, a lepton isodoublet $(E^+_2,N_1^0)_L^T$ with
($N_{1L}^{0c}=N_{2L}^0$ in the weak basis), and two charged lepton
isosinglets $E^+_{1L}$ and $E^+_{3L}$. Besides, a right-handed neutrino field is naturally included in the weak basis.

The model in Ref.\cite{spm} is also a one family model like the one presented here and uses the same value for the $b$ parameter in the electric charge generator $Q$; so, the gauge boson content of both models is the same. Now, for the model in Ref.\cite{spm} there are 27 Weyl fields which are the same fields present in the fundamental representation of the electroweak-strong unification group $E_6$ \cite{e6}; in our model there are 30 Weyl fields which can be accommodated in the representations of a new electroweak-strong unification group $SU(6)\otimes U(1)_X$ which is not a subgroup of $E_6$, so the fermion content of the two models is different.

The low energy predictions of the two models are also different. 
Even though the low energy charged sector of both models looks alike, it 
is not so for the neutral sector. The model in Ref.\cite{spm} allows for a second neutral current associated with a mass as low as 600 GeV and predicts for each family the existence 
of a tiny mass Majorana neutrino and a Dirac neutrino with a mass of the order of the electroweak mass scale. For the model in this paper there is only a small mass Dirac neutrino in each family and the mass scale of the second neutral current is of the order of a few TeV. Also the scalar sector of the two models is quite different.

Finally we like to emphasize the way how the fermion 
masses in one family are generated in the model presented in this 
paper:
First, all the exotic fermions get very large masses, while the lightest charged lepton acquires a mass at tree level of the order of $(h_1 - h_3)v$ (in the limit $V \gg v$), suppressed as a consequence of the mixing with exotic charged leptons. The up quark gets a mass at tree level suppressed also as a consequence of a mixing with a heavy exotic up type quark. The down quark gets its mass as a one-loop radiative correction. Finally there is a Dirac neutrino in each family with a mass suppressed by differences of 
products of Yukawa couplings. This unfamiliar way to generate masses 
might lead to explanations for the major trends in the masses of the 
fermions within one family (the first family for example).

\section{ACKNOWLEDGMENTS}
This work was partially supported by BID and Colciencias in Colombia.

\section*{APPENDIX}

In this appendix we diagonalize perturbatively the mass matrix $M_{eE}$ for the charged leptons which appears in section 5.2.2. This matrix can be written as

\[M_{eE}=V\left(\begin{array}{cccc} h_{11}q & h_{12}q &
- h_{1} & h_{13}q\\ h_{5} & h_{6} &  h_{2}q & h_{7}\\ h_{14}q & h_{15}q &
- h_{3} & h_{16}q\\ h_8  & h_9  &  h_{4}q & h_{10} \end{array} \right),\]

\noindent where $q=v/(\sqrt{2}V)\ll 1$.\\

The matrix $M^2(v,V)=M_{eE}M^T_{eE}$ can be separated as $M^2=V^2(M^2_0 + M^2(q))$, where $V^2M^2_0 = M^2(v=0,V)$ and

\[M^2_0=\left(\begin{array}{cccc} h^2_1 & 0 &
h_1h_3 & 0\\ 0 &  A^2 & 0 & B^2\\ h_1h_3 & 0 &
h^2_3 & 0\\ 0 & B^2 &  0 & C^2 \end{array} \right),\]

\[M^2(q)=\left(\begin{array}{cccc} J^2q^2 & D^2q &
K^2q^2 & E^2q\\ D^2q & h^2_2q^2 &  F^2q & h_2h_4q^2\\ K^2q^2 & F^2q &
H^2q^2 & G^2q\\ E^2q  & h_2h_4q^2  &  G^2q & h^2_4q^2 \end{array} \right).\]

In $M^2_0$ and $M^2(q)$ the parameters are related to the Yukawas in the following way

\begin{eqnarray} \nonumber
A^2&=&h^2_5+h^2_6+h^2_7\;, \\ \nonumber
B^2&=&h_5h_8+h_6h_9+h_7h_{10}\;,\\ \nonumber
C^2&=&h^2_8+h^2_9+h^2_{10}\;, \\ \nonumber
D^2&=&-h_1h_2+h_5h_{11}+h_6h_{12}+h_7h_{13}\;,\\ \nonumber
E^2&=&-h_1h_4+h_8h_{11}+h_9h_{12}+h_{10}h_{13}\;,\\ \nonumber
F^2&=&-h_2h_3+h_5h_{14}+h_6h_{15}+h_7h_{16}\;,\\ \nonumber
G^2&=&-h_3h_4+h_8h_{14}+h_9h_{15}+h_{10}h_{16}\;,\\ \nonumber
H^2&=&h^2_{14}+h^2_{15}+h^2_{16}\;, \\ \nonumber
J^2&=&h^2_{11}+h^2_{12}+h^2_{13}\;, \\ \nonumber
K^2&=&h_{11}h_{14}+h_{12}h_{15}+h_{13}h_{16}\;.\\ \nonumber
\end{eqnarray}

The eigenvalues of the matrix $V^2M^2_0$ are
\begin{eqnarray} \nonumber
m^2_{01}&=&0\;, \\ \nonumber
m^2_{02}&=&(h^2_1+h^2_3)V^2\;,\\ \nonumber
m^2_{03}&=&(A^2+C^2-\sqrt{A^4+4B^4+C^4-2A^2C^2})V^2/2\;, \\ \nonumber
m^2_{04}&=&(A^2+C^2+\sqrt{A^4+4B^4+C^4-2A^2C^2})V^2/2\;.\\ \nonumber
\end{eqnarray}

We now rotate $M^2(q)$ using the matrix of eigenvectors of $M^2_0$. After the algebra is done we get, at first order in the expansion $v^2/V^2$, the following four eigenvalues

\begin{eqnarray} \nonumber
m^2_1&=&[(h_1^2H^2 + h_3^2 J^2 - 2h_1h_3K^2)/[2(h_1^2 + h_3^2)]] v^2\;, \\ \nonumber
m^2_2&=&m^2_{02}+O(v^2)\;,\\ \nonumber
m^2_3&=&m^2_{03}+O(v^2)\;, \\ \nonumber
m^2_4&=&m^2_{04}+O(v^2)\;,\\ \nonumber
\end{eqnarray}

\noindent where $m_1$ corresponds to the mass of the lightest charged lepton and $O(v^2)$ stands for corrections of order $v^2$ to the squared masses of the exotic charged leptons.

\newpage
\vspace{3.0cm}
\begin{figure}[t]
\epsfxsize=12 truecm \epsfysize=12 truecm \centerline{\epsffile[20 20 500 550]{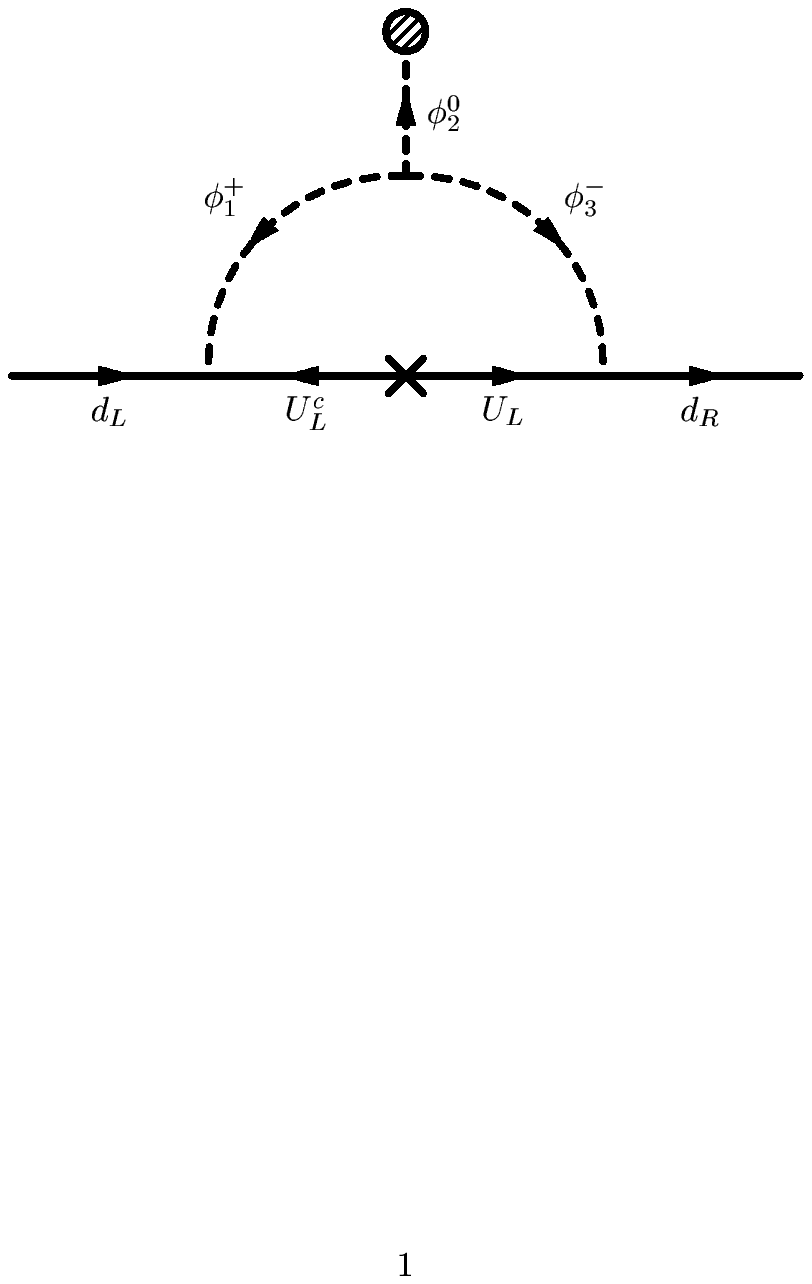}}
\end{figure}

\noindent Figure 1: A one-loop diagram contributing to the radiative generation of the down quark mass.

\newpage

\begin{figure}[t]
\epsfxsize=10 truecm \epsfysize=12 truecm \centerline{\epsffile[20 22 500 550]{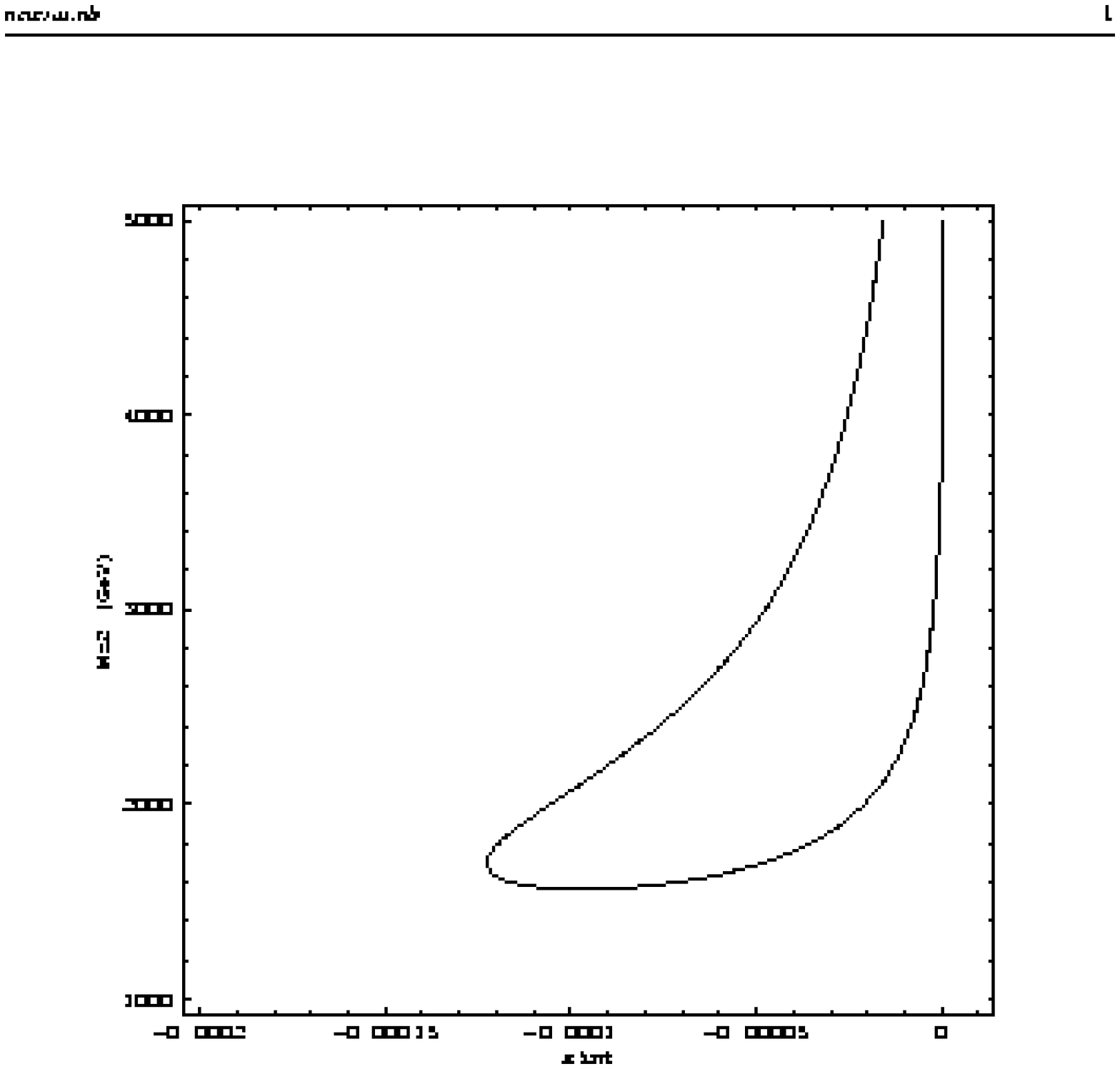}}
\end{figure}

\noindent Figure 2: Contour plot displaying the allowed region for $\sin\theta$ vs. $M_{Z_2}$ at 95$\%$ CL.


\newpage


\begin{thebibliography}{99}
\bibitem{sm}
S.L. Glashow, Nucl. Phys. {\bf 22}, 579 (1961); S. Weinberg, Phys. Rev. Lett. {\bf 19}, 1264 (1967); A. Salam, in {\it Elementary Particle Theory: Relativistic Groups and Analyticity (Nobel Symposium No.8)}, edited by N. Svartholm (Almqvist and Wiksell, Stockholm, 1968), p.367

\bibitem{su3}
For a review see: W. Marciano and H. Pagels, Phys. Rep. {\bf 36C}, 137
(1978).

\bibitem{ext}
For discussions and reviews, see: R.N. Mohapatra, 
{\it Unification and Supersymmetry}
(Springer, New York, 1986); P. Langacker, Phys. Rep. {\bf C72}, 
185 (1981); H. E. Haber and G. L. Kane, Phys. Rep. 
{\bf 117}, 75 (1985); M. B. Green, J. H. Schwarz and E. Witten, 
{\it Superstring Theory, Vols. 1 \& 2} (Cambridge University Press, 
Cambridge, 1987).

\bibitem{lee}
G. Segr\`e and J. Weyers, Phys. Lett. {\bf B65}, 243 (1976); B.W. Lee
and S. Weinberg, Phys. Rev. Lett. {\bf 38},1237 (1977); P. Langacker
and G. Segr\`e, Phys. Rev. Lett. {\bf 39}, 259 (1977); M. Singer,
Phys. Rev. {\bf D19}, 296 (1979); K.T. Mahanthappa and P.K. Mohapatra, 
Phys. Rev. {\bf D42}, 1732 (1990); 
{\bf D42}, 2400 (1990); {\bf D43}, 3093 (1991).

\bibitem{fp}
F. Pisano and V. Pleitez, Phys. Rev. {\bf D46}, 410 (1992);  P.H. Frampton,
Phys. Rev. Lett. {\bf 69}, 2887 (1992); J.C. Montero, 
F. Pisano and V. Pleitez, Phys. Rev. {\bf D47}, 2918 
(1993); R. Foot, O.F. Hernandez,
F. Pisano and V. Pleitez, Phys. Rev. {\bf D47}, 4158 (1993); 
V. Pleitez and M.D. Tonasse, Phys. Rev. {\bf D48}, 2353 (1993); {\it ibid} 5274 (1993); D. Ng, Phys. Rev. {\bf D49}, 4805 (1994); L. Epele, H. Fanchiotti, C. Garc\'\i a Canal and D. G\'omez Dumm, Phys. Lett. {\bf B343} 291 (1995); M. \"Ozer, Phys. Rev.{\bf D54}, 4561 (1996); D. G\'omez Dumm, Phys. Lett. {\bf B411},313 (1997).

\bibitem{long}
M. Singer, J.W.F. Valle and J. Schechter, Phys. Rev. {\bf D22}, 738 (1980); 
R. Foot, H.N. Long and T.A. Tran, Phys. Rev. {\bf D50}, R34 (1994); H.N.
Long, Phys. Rev. {\bf D53}, 437 (1996); {\it ibid} {\bf D54}, 4691 (1996);
V. Pleitez, Phys. Rev. {\bf D53}, 514 (1996).

\bibitem{ponce}
W.A. Ponce, J.B. Fl\'orez and L.A. S\'anchez,  
{\it "Analysis of the $SU(3)_c\otimes SU(3)_L\otimes U(1)_X$ 
local gauge theory"}, hep-ph/0103100, Int. J. Mod. Phys. {\bf A} (to be published).

\bibitem{spm}
L.A. S\'anchez, W.A. Ponce and R. Mart\'\i nez, Phys. Rev. {\bf D64}, 075013 (2001).

\bibitem{del}
R. Delbourgo, Phys. Lett. {\bf B40}, 381 (1972); L. Alvarez-Gaume, 
Nucl. Phys., {\bf B234}, 262 (1984).

\bibitem{georgi}
H. Georgi and S.L. Glashow, Phys. Rev. {\bf D6}, 429 (1972); 
A. Fern\'andez and R. Mart\'\i nez, Rev. Mex. F\'\i s., 
{\bf 35} No.3, 379 (1989).

\bibitem{slansky}
R. Slansky, Phys. Rep. {\bf 79}, 1 (1981).

\bibitem{flipped}
S. Cecotti {\it et al}, Phys. Lett. {\bf B156}, 318 (1985); G. Lazarides and Q. Shafi, Nucl. Phys. {\bf B329}, 183 (1990); {\bf B338}, 442 (1990); C. Panagiotakopoulos, Int. J. Mod. Phys. {\bf A5}, 2359 (1990).

\bibitem{schiff}
L.I. Schiff. {\it Quantum Mechanics} (McGraw-Hill, New York, 3rd Ed., 1968), pp. 248-250.

\bibitem{lang2001}
S.C. Bennett and C.E. Wieman, Phys. Rev. Lett. {\bf 82},
2484 (1999); Paul Langacker, hep-ph/0102085.

\bibitem{data} 
Particle Data Group, D.E. Groom {\it et. al.}, Eur. Phys. J.
{\bf C15}, 1 (2000); G. Altarelli, hep-ph/0011078.

\bibitem{pich} 
J. Bernabeu, A. Pinch and A. Santamaria, Nucl. Phys.
{\bf B363}, 326 (1991).


\bibitem{marciano}
W.J. Marciano and A. Sirlin, Phys. Rev. {\bf D29}, 75
(1984); W. Marciano and J.L. Rosner, Phys. Rev: Lett. {\bf 65}, 2963
(1990).

\bibitem{durkin} 
L. Durkin and P. Langacker, Phys. Lett. {\bf B166}, 436
(1986).

\bibitem{todos}
R. Casalbuoni, S. De Curtis, D. Dominici and R. Gatto,
Phys. Lett. {\bf B460}, 135 (1999); J.L. Rosner, Phys. Rev. {\bf D61},
016006 (2000); J. Erler and P. Langacker, Phys. Rev. Lett. {\bf 84}, 212
(2000).

\bibitem{fermi}
F. Abe {\it et. al.}, Phys. Rev. Lett. {\bf 79}, 2192 (1997).

\bibitem{e6}
F. G\"ursey, P. Ramond and P. Sikivie, Phys. Lett. {\bf B60}, 177 (1975); 
F. G\"ursey and M. Serdaroglu, Lett. Nuovo Cimento Soc. Ital. Fis. 
{\bf 21}, 28 (1978).

\end{thebibliography}
\end{document}